# Highly resilient, error-protected quantum gates in a solid-state quantum network node

E. Poem-Kalogerakis,[1] M. I. Cohen,[1] S. Blum,[1] D. Minin,[1] D. Korn,[1] O. Heifler,[1] A. Hamo,[1] I. Bayn,[1] N. Bar-Gill,[1,2] R. L. Patel,[3] M. Markham,[3] and M. Tordjman[1]

[1] *Quantum Transistors Ltd., Binyamina 3057319, Israel*
[2] *The Racah Institute of Physics, The Hebrew University, Jerusalem 9190401, Israel*
[3] *Element Six Global Innovation Centre, Didcot, OX11 0QR, United Kingdom.*

High-fidelity quantum gates are the cornerstone of any quantum computing and communications architecture. Realizing such control in the presence of realistic errors at the level required for beyond-threshold quantum error correction is a long-standing challenge for all quantum hardware platforms. Here we theoretically develop and experimentally demonstrate error-protected quantum gates in a solid-state quantum network node. Our work combines room-temperature randomized benchmarking with a new class of composite pulses that are simultaneously robust to frequency and amplitude, affecting random and systematic errors. We introduce Power-Unaffected, Doubly-Detuning-INsensitive Gates (PUDDINGs) - a theoretical framework for constructing conditional gates with immunity to both amplitude and frequency errors. For single-qubit and two-qubit gate demonstrations in a solid-state nitrogen-vacancy (NV) center in diamond, we tested the operation of the resulting protected gates at room temperature and find improvements above an order of magnitude. For a two-qubit CNOT gate, between the nuclear and electron spins, an error per gate of $2.7 \pm 1.9 \times 10^{-4}$ is measured. Taking into account the increase in the electron coherence time at low temperatures, we predict that a two-qubit error per gate around $3 \times 10^{-5}$ is achievable. These results, far below the thresholds required by surface and color code error correction, thus present viable building blocks for a new class of fault-tolerant quantum networks.

## I. INTRODUCTION

Fault-tolerant quantum computation requires gate fidelities that exceed stringent error-correction thresholds, typically in the range of $10^{-4}$ to $10^{-2}$ for leading quantum error-correcting codes such as surface codes and color codes [1]. Achieving and sustaining such performance in realistic devices is challenging because gate operations must be robust not only to static calibration errors but also to time-dependent noise and drift in both the control fields and the surrounding environment [2].

Nitrogen-vacancy (NV) centers in diamond [3] offer unique advantages for distributed quantum networks. The NV center's optical addressability, long coherence times, and intrinsic coupling to nearby nuclear spins make it an attractive candidate for quantum communication and distributed quantum computing nodes. Recent studies on NV centers have demonstrated promising single-qubit operation fidelities [4], but conditional two-qubit gates fidelities, essential for universal quantum computation and entanglement-based networking, remain limited [5]. This is mainly due to the sensitivity of these gates to spectral diffusion, control-power fluctuations, and dephasing. Therefore, a reliable gate protection strategy is of imminent importance.

Dynamical decoupling and composite-pulse techniques offer complementary routes to error suppression at the control level. Dynamical decoupling sequences [2], originally developed in nuclear magnetic resonance, can strongly suppress quasi-static dephasing during idle periods, while composite pulses [6] engineer gate operations from carefully phased sub pulses to cancel systematic amplitude and detuning errors to first or even higher order. While some broad-band composite pulses (BB-COMPs) that protect against both frequency and amplitude errors are known [7], they apply only for single-qubit $\pi$-rotations, and no BB-COMPs are known for two-qubit, conditional gates.

Here we first find a generalized family of BB-COMPs that protect arbitrary-angle single-qubit rotations against both amplitude and frequency errors. Then we introduce a new class of amplitude and frequency error-protected conditional, two-qubit gates, which we call Power-Unaffected, Doubly-Detuning-INsensitive Gates (PUDDINGs). Finally, we experimentally perform these gates on single NV centers in isotopically-purified diamond at room temperature, and test their performance with respect to unprotected gates. Single-qubit gates are applied to the nuclear spin of the nitrogen atom, and we use randomized benchmarking (RB) [8] on the full Clifford group for measuring the average error per gate (EPG). A two-qubit CNOT gate is applied to the electron spin conditioned on the nuclear spin, and interleaved randomized benchmarking (IRB) [9] is used to measure its EPG.

As our experiments are performed at room temperature, the EPG is ultimately limited by the electron spin lifetime ($T_1$) of a few milliseconds [10]. This limitation is significantly alleviated at low temperatures, where the electron spin lifetime exceeds seconds. This allows us to use our room-temperature results to put an upper bound on the expected low-temperature EPGs by subtracting the (independently measured) $T_1$ contribution. For single-qubit gates this results in an EPG of $8.6 \pm 6 \times 10^{-5}$. Compared to the EPG of the unprotected gates, $1.8 \pm 0.2 \times 10^{-3}$, this constitutes a 20-fold improvement. For the PUDDING CNOT gate, we measure a $T_1$-subtracted EPG of $2.7 \pm 1.9 \times 10^{-4}$. Compared to the $5.1 \pm 0.2 \times 10^{-3}$ value measured for the unprotected gate, this is again an improvement of about 20-fold. Using numerical simulations of dynamical noise we show that our measured PUDDING CNOT EPG is limited by a combination of the coherence times $T_2$ and $T_2^*$ of the electron spin (independantly measured to be $700\,\mu s$ and $50\,\mu s$, respectively) and pulse amplitude fluctuations of a few tenths of a percent. Considering ultra-high isotopic purity diamond at low temperature ($T_2 \approx 1800\,\mu s$ [4] and $T_2^* \approx 400\,\mu s$ [11]) and relative amplitude fluctuations of $10^{-3}$ [12], we project that an error probability of about $3 \times 10^{-5}$ is achievable.

The remainder of this manuscript is organized as follows. In Sec. II we develop the theoretical framework for error-protected conditional quantum gates. Sec. III describes the experimental NV platform, and in Sec. IV we present the experimental results, comparing unprotected and PUDDING-protected single- and two-qubit gates at room temperature and extracting projected performance at low temperature in isotopically purified diamond. Sec. V discusses the implications of these results for fault-tolerant quantum computation and quantum networking and compares our projected performance to that of other leading platforms. Finally, Sec. VI concludes with an outlook on future applications and extensions of PUDDINGs to other qubit systems. Technical derivations and details are provided in the Appendices.

## II. THEORY

### A. Error principles in Quantum Operations

Consider a target single-qubit rotation

$$U_{target} = e^{-i\alpha\hat{n}\cdot\vec{\sigma}/2} = \cos\tfrac{\alpha}{2} I - i\sin\tfrac{\alpha}{2}\hat{n}\cdot\vec{\sigma}, \tag{1}$$

where $I$ is the unit matrix, $\vec{\sigma} = (\sigma_x, \sigma_y, \sigma_z)$ is a vector of Pauli matrices, $\alpha$ is the rotation angle, and $\hat{n} = (\cos\theta\cos\phi, \cos\theta\sin\phi, \sin\theta)$ is the rotation axis. Here $\phi$ is the drive phase, $\tan\theta = \delta/\Omega$, and $\alpha = T\sqrt{\Omega^2 + \Delta^2}$, where $\Omega$ is the Rabi frequency, proportional to the pulse amplitude, $\delta$ is the detuning from resonance, and $T$ is the pulse duration. Typically, resonant driving is used, so $\delta = \theta = 0$, such that

$$U_{target} = \cos\tfrac{\alpha}{2} I - i\sin\tfrac{\alpha}{2}[\cos\phi\sigma_x + \sin\phi\sigma_y]. \tag{2}$$

As illustrated in Fig. 1(a), real pulses suffer from systematic and random errors,

$$\Omega \to \Omega(1 + \epsilon(t)), \tag{3}$$
$$\delta \equiv \Omega\eta(t) \neq 0. \tag{4}$$

In most cases, timing and phase errors can be neglected with respect to amplitude and detuning errors. Further, for small systematic or quasi-static errors, $\epsilon(t) = \epsilon \ll 1$, $\eta(t) = \eta \ll 1$, when the noise is approximately constant over the pulse duration, one obtains that to the lowest order in the noise magnitudes, the difference between the actual operation and the target operation is

$$\begin{aligned} U_{actual} - U_{target} \approx& -\epsilon\tfrac{\alpha}{4}\sin\tfrac{\alpha}{2} I \\ &- i\epsilon\tfrac{\alpha}{4}\cos\tfrac{\alpha}{2}(\cos\phi\sigma_x + \sin\phi\sigma_y) - i\eta\sin\tfrac{\alpha}{2}\sigma_z. \end{aligned} \tag{5}$$

The error per gate (EPG) can then be quantified by using the Frobenius distance of $U_{actual}$ from $U_{target}$. This is just the square root of the mean of the squares of the differences between the elements of the matrices,

$$\begin{aligned} EPG(\epsilon,\eta) &\approx ||U_{actual} - U_{target}||_F \\ &= \frac{1}{\sqrt{2}}\sqrt{\tfrac{\alpha^2}{16}\epsilon^2 + \sin^2\tfrac{\alpha}{2}\eta^2}. \end{aligned} \tag{6}$$

For this case of a simple resonant pulse, the EPG is thus linear in some combined magnitude of the pulse amplitude ($\epsilon$) and frequency ($\eta$) noise. Looking at Eq. (5), we further see that, for this case, amplitude noise causes bit-flip errors ($\sigma_x$ and $\sigma_y$), while frequency noise causes phase errors ($\sigma_z$). The effects of these different error types on a qubit are illustrated in Fig. 1(b). When taking the average EPG of many error values sampled from Gaussian distributions, $G_{w_\epsilon}(\epsilon)$ and $G_{w_\eta}(\eta)$, of widths (standard deviations) $w_\epsilon$ and $w_\eta$ for the amplitude and frequency errors, respectively, one obtains

$$\begin{aligned} EPG &= \sqrt{\int_{-\infty}^{\infty} EPG^2(\epsilon,\eta) G_{w_\epsilon}(\epsilon) G_{w_\eta}(\eta) d\epsilon d\eta} \\ &\approx \frac{1}{\sqrt{2}}\sqrt{\tfrac{\alpha^2}{16}w_\epsilon^2 + \sin^2\tfrac{\alpha}{2}w_\eta^2}. \end{aligned} \tag{7}$$

The inhomogeneous spectral width $\Omega w_\eta$ is Fourier related to the inhomogeneous dephasing ($1/e$) time, $T_2^*$, by $w_\eta = \sqrt{2}/\Omega T_2^*$. Equivalently, one can define an inhomogeneous flip time, $T_1^*$, such that $w_\epsilon = \sqrt{2}/\Omega T_1^*$. As the gate duration is $T_g = \alpha/\Omega$, we have

$$EPG \approx \frac{1}{2}\sqrt{\frac{1}{4}\left(\frac{T_g}{T_1^*}\right)^2 + \mathrm{sinc}^2\frac{\alpha}{2}\left(\frac{T_g}{T_2^*}\right)^2}. \quad (8)$$

Particularly, when the dominant noise source is frequency noise, $EPG \propto T_g/T_2^*$.

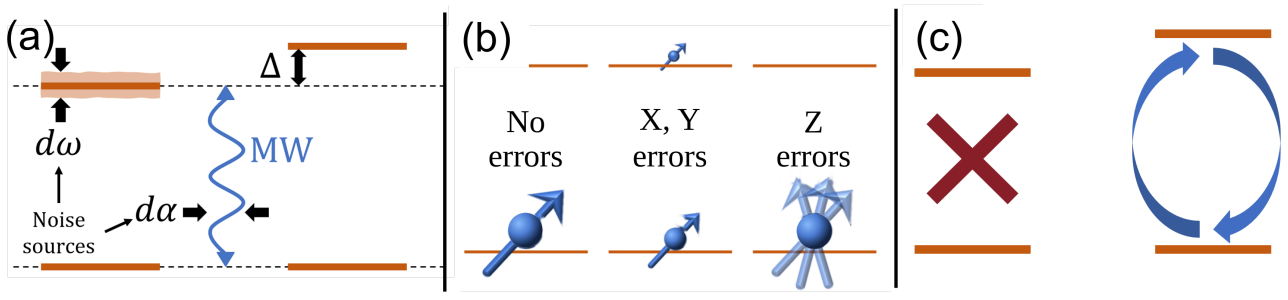

| (d) Gate Type | Noise Source | Conditioning Qubit = 0 | | Conditioning Qubit = 1 | |
|---|---|---|---|---|---|
| | | X, Y Errors | Z Errors | X, Y Errors | Z Errors |
| Unprotected | $d\alpha$ | ✗ | ✗ | ✗ | ✗ |
| | $d\omega$ | ✗ | ✗ | ✗ | ✗ |
| DD, P - ZAP | $d\alpha$ | ✓ | ✓ | ✗ | ✗ |
| | $d\omega$ | ✓ | ✓ | ✗ | ✗ |
| PUDDING | $d\alpha$ | ✓ | ✓ | ✓ | ✓ |
| | $d\omega$ | ✓ | ✓ | ✓ | ✓ |

FIG. 1. Noise sources and error types. (a) Noise sources – $d\alpha$ is pulse amplitude (Rabi frequency) noise and $d\omega$ is resonance frequency noise. (b) Error types – X and Y errors change state populations, while Z errors change the relative phase between the states. (c) Conditional two-qubit gate. two of the four levels are acted upon (the 'active' subspace') while the other two are not (the 'inactive' subspace). (d) The ability of different gate protocols to handle errors from different noise channels. Unprotected gates relate to simple ZAPs (see text). DD protocols are sequences based on Dynamical Decoupling [2–4, 13–18]. PUDDING protocol is the main protocol discussed in this work.

Dynamical decoupling [19, 20] can suppress dephasing during idle periods by periodically flipping the qubit phase, but it cannot protect the quantum state during gate operation where the desired rotation must occur. Composite pulses [21–27] address this by constructing the target operation from sub-pulses with carefully chosen phases and amplitudes such that error terms cancel to first (or higher) order. In a similar way to the derivation above, one can show that for a composite pulse with $n^{th}$-order protection (cancellation of all error terms up to and including the $n^{th}$ order), the EPG due to frequency noise (when neglecting amplitude noise), would scale as $(T_g/T_2^*)^{n+1}$.

For a two-qubit CNOT gate, as shown in Fig. 1(c), one can split the operation to two two-level subspaces. In one, the 'active' subspace, the gate operation is a $\pi$ rotation, while in the other, 'inactive' subspace, it is the unit operation. When cross-talk errors, mixing between the two subspaces, are negligible, which is the case when, for example, cross-transitions are forbidden by symmetry, the total gate error is just the quadratic sum of the errors in the two sub-spaces. Thus, to suppress errors in a CNOT gate, one would have to simultaneously suppress errors in both the 'active' and 'inactive' subspaces. Figure 1(d) qualitatively summarizes the level of protection offered by existing two qubit gate schemes, including unprotected two-qubit gates, dynamical decoupling protocols (CPMG [4, 13, 14], XY8 [3, 15–17], AXY [2, 18]), as well as the PUDDING gates introduced below. For each gate realization and for each part of the gate, we indicate whether or not it is vulnerable to either frequency or amplitude noise, and whether the vulnerability is expressed in bit-flip or phase errors. PUDDING gates are the only strategy that protects all gate parts and possible error types against both amplitude and detuning noise.

## B. Broad-band composite pulses (BB-COMPs)

Broad-band composite pulses (BB-COMPs) [21–27] compensate systematic errors and quasi-static noise by nullifying derivatives of the EPG with respect to noise parameters. A composite $\pi$-pulse comprises of N simple $\pi$-pulses with phases $\phi_k$:

$$U_{\pi,comp} = \prod_{k=1}^{N} U_{\pi,\phi_k}, \quad (9)$$

where

$$U_{\pi,\phi_k} = e^{-i\pi[\cos\phi_k\sigma_x + \sin\phi_k\sigma_y]/2}. \quad (10)$$

For a sequence of three $\pi$ pulses with phases of $\{0,\phi,0\}$, choosing $\phi = \pi/2$ nullifies the first-order amplitude derivative of the EPG, $\partial EPG/\partial\epsilon|_{\epsilon=0} = 0$. However, three pulses of equal rotation angle cannot simultaneously nullify both $\partial EPG/\partial\epsilon$ and $\partial EPG/\partial\eta$. Five-pulse sequences with optimized phases achieve such dual protection. The U5a sequence [23, 27, 28] employs five $\pi$ pulses with phases of $\{0, 5\pi/6, \pi/3, 5\pi/6, 0\}$, achieving $\partial EPG/\partial\epsilon|_{\epsilon=\eta=0} = \partial EPG/\partial\eta|_{\epsilon=\eta=0} = 0$ with a total duration $T_{U5a} = 5\pi/\Omega$. Such composite pulses share the property that second-order derivatives are non-zero, and thus residual errors scale as $O(\epsilon^2, \eta^2, \epsilon\eta)$. This quadratic scaling means that for a given inhomogeneous dephasing time $T_2^*$, the error in a gate of length $T_g$ scales as $(T_g/T_2^*)^2$ instead of $T_g/T_2^*$ as for simple, unprotected gates, as discussed above. This is the reason why composite pulses may provide an advantage even though they may be considerably longer than simple pulses.

Protected composite pulses that rotate by angles other than $\pi$ are known [14, 29], but with protection only against amplitude errors. To include protection also against detuning errors for any desired rotation angle, $\alpha$, we consider a seven-pulse sequence made of five $\pi$-pulses sandwiched between two $(\pi-\alpha)/2$ pulses, and found the required phases to be (Appendix A)

$$\{\phi\} = \{0, h, 2h-g, 2h-2g+\pi, 2h-g, h, 0\}, \quad (11)$$

where

$$g(\alpha) = \cos^{-1}\left(-\frac{1}{2}+\frac{\alpha}{4\pi}+\frac{\sin\frac{\alpha}{2}}{4}\right), \quad (12)$$

$$h(\alpha) = \cos^{-1}\left(\frac{1}{2}+\frac{\alpha}{4\pi}-\frac{\sin\frac{\alpha}{2}}{4}\right). \quad (13)$$

For $\alpha=\pi$, this solution reduces to U5a (U5b) if one takes the negative (positive) branch of the arc cosine. This generalization allowed us to achieve the full Clifford group RB of protected single qubit gates (see Sec. III below).

### C. Energy-selective conditional gates

Consider a two-qubit system with states $\{|00\rangle, |01\rangle, |10\rangle, |11\rangle\}$,where qubit $q_1$ (control) modulates qubit $q_2$'s (target) transition frequency by $\Delta$:

$$\omega_2^{(q_1=0)} = \omega_0, \quad (14)$$

$$\omega_2^{(q_1=1)} = \omega_0+\Delta. \quad (15)$$

The goal is to flip $q_2$ conditionally: $|q_1q_2\rangle \to |q_1\overline{q_2}\rangle$ only for $q_1=1$, while for $q_1=0$ leaving $q_2$ unchanged. There are several known ways of achieving this.

*1. Weak-Interaction Regime ($\Omega \ll \Delta$)* This is the simplest and most commonly used method [4, 13, 30, 31]. Tune a resonant pulse to $\omega_0+\Delta$. Off-resonant excitation of the $\omega_0$ transition is then suppressed. However, it is never exactly zero, even without any noise, always leaving a residual gate infidelity of $|\Omega/\Delta|$. As the required gate time is $\pi/\Omega$, decreasing the infidelity requires increasing the pulse duration, eventually running into the inhomogeneous dephasing time, $T_2^*$. This limits the minimal gate error to $\sim \pi/\Delta T_2^*$.

*2. Strong-Interaction Regime ($\Omega \gg \Delta$)* Apply two pulses of duration $T_p=\pi/2\Omega$ and center frequency $\omega_0+\Delta/2$, phase-shifted by $\pi/2$ and separated by $T_{sep}=\pi/\Delta$ [32]. This induces a $\pi$ rotation for the $\omega_0+\Delta$ transition while the $\omega_0$ transition acquires only a phase. The residual error, even without noise, scales as $|\Delta/\Omega|$, requiring large $\Omega$, which is not always available.

*3. $\pi-2\pi$ Gate ($\Omega\sim\Delta$)* This is an exact conditional gate. It applies a single pulse, resonant with $\omega_0+\Delta$, choosing $\Omega=\Delta/\sqrt{3}$ and $T_P=\sqrt{3}\pi/\Delta$. This choice ensures that the resonant transition undergoes a $\pi$ rotation, while the detuned transition experiences a $2\pi$ rotation, returning to its initial state [33]. While exact, this gate is still first-order sensitive to both $\epsilon$ and $\eta$ for both transitions (as can be seen from Eq. 8).

Unfortunately, none of these approaches protect either of the inactive (unity operation) or the active (flip operation) subspace against systematic or random errors. This motivates our development of P-ZAPs and PUDDINGs.

### D. Protected zero-area pulses (P-ZAPs)

A zero-area pulse (ZAP) satisfies the condition

$$\int_{\infty}^{\infty}\Omega(t)dt = 0. \quad (16)$$

Such pulses cannot affect a two-level system which frequency is at the mean frequency of the pulse. However, they can affect detuned two level systems [34]. This makes ZAPs inherently energy-selective, and therefore useful for implementing energy-selective conditional gates.

The simplest ZAP consists of two rectangular pulses of equal amplitude and duration but opposite phase. For a resonant two-level system, the first pulse rotates the Bloch vector away from the north pole and the second pulse reverses this rotation, yielding a net identity operation. Any constant amplitude error scales both halves equally, preserving the zero net area and hence the identity to all orders in the amplitude error. For a detuned transition with detuning $\delta=\Delta$, choosing $\Omega=\Delta$ and $T_p=\pi/\sqrt{2}\Delta$ leads to a net $\pi$ rotation of the detuned transition while leaving the resonant transition as identity. Thus, this ZAP acts as a simple conditional gate: identity on resonance and a $\pi$ flip at detuning of $\Delta$ from resonance. However, it remains first-order sensitive to detuning errors on the resonant transition and to both detuning and amplitude errors on the detuned transition. Fig. 2(a) presents this pulse and its effects on the Bloch spheres of the on-resonance and detuned subspaces. Residual first-order errors remain in all the components of the over-all rotation.

To suppress first-order detuning errors on the resonant transition, we construct time-symmetric ZAPs. Using the Magnus expansion, one can show (Appendix B 1) that if a ZAP is symmetric (for example, ABBA structure with A and B equal-amplitude pulses of opposite phases), the first order $\sigma_x$ and $\sigma_y$ error terms vanish. For a $\theta$ rotation on the detuned transition we find (Appendix C) that the ABBA sequence has to have

$$\Omega_A = -\Omega_B = \Delta\sin\frac{\theta}{4}, \quad (17)$$

$$\alpha_A = \alpha_B = \frac{\sin\frac{\theta}{4}}{\sqrt{\sin^2\frac{\theta}{4}+1}}\cos^{-1}\left(-\sin^2\frac{\theta}{4}\right). \quad (18)$$

In particular, for $\theta=\pi$,

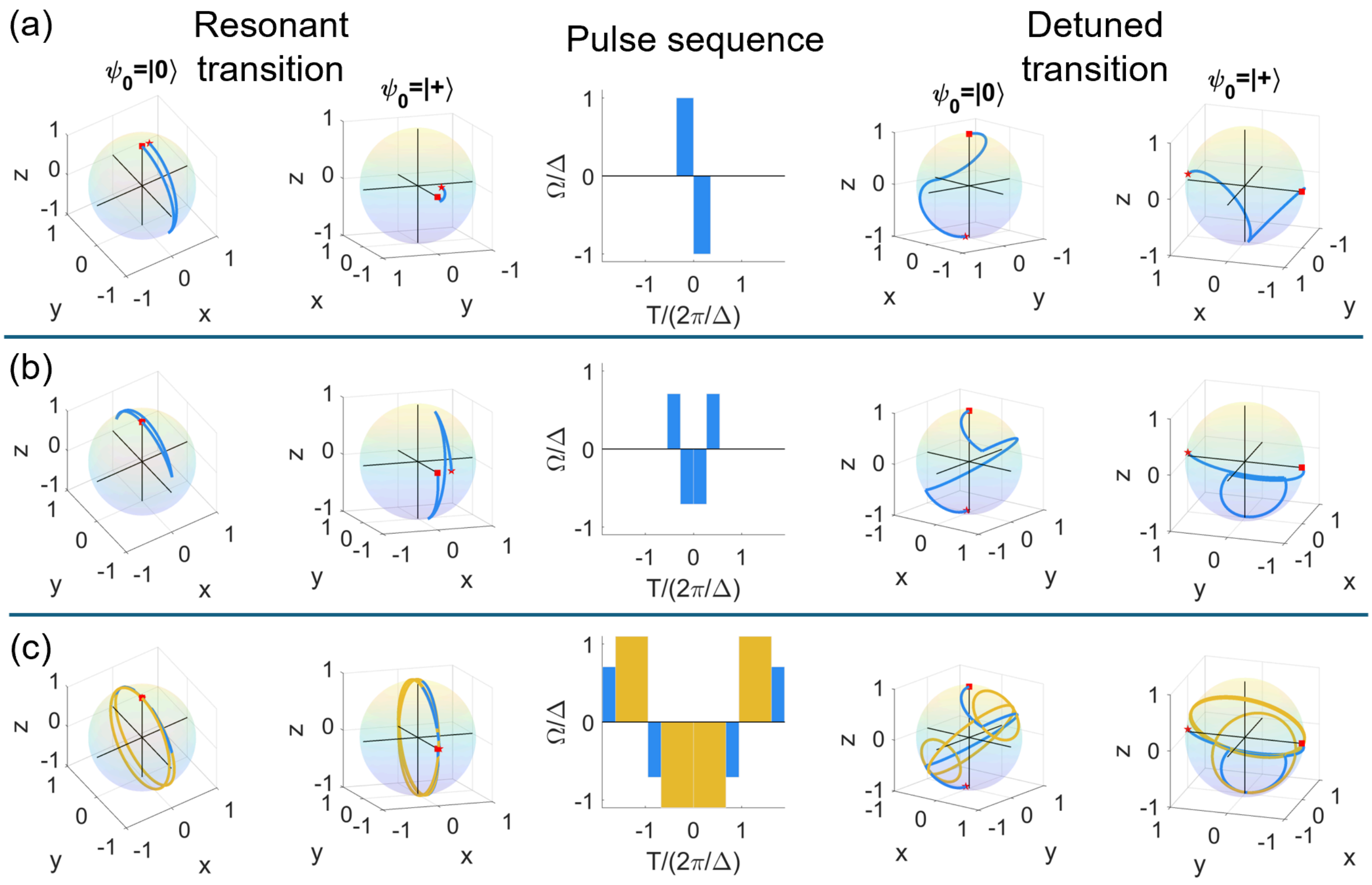

FIG. 2. Zero-area conditional gate pulse sequences. (a) The simple ZAP gate. (b) The symmetric ZAP gate. (c) The protected ZAP (P-ZAP) gate (Eq. (23) and Eq. (24), $p = +1$). In each row, the middle panel presents the Rabi frequency profile, and the left (right) two panels present Bloch sphere trajectories for the on-resonance ($\Delta$-detuned) transition for two different initial states (red squares), for a detuning error of $0.05\Delta$. The final states are marked by red stars. The augmenting pulses of the P-ZAP are colored yellow. One can see that for the on-resonance transition, the P-ZAP (c) gate is the only one that compensates for the detuning error for both initial states.

$$\Omega_A = -\Omega_B = \Delta/\sqrt{2}, \tag{19}$$

$$\alpha_A = \alpha_B = 2\pi/\sqrt{27}. \tag{20}$$

Fig. 2(b) presents this pulse and its effects on the Bloch spheres of the on-resonance and detuned subspaces. For the former, first-order errors remain only in the $\sigma_z$ component.

Complete first-order protection of the identity operation on the resonant transition requires another free parameter to control. We thus augment each component of the symmetric ZAP with a compensating pulse. Consider the sequence ACBDDBCA, where A, B are the original opposite-sign rotations and C, D are augmenting pulses designed to cancel the on-resonance $\sigma_z$ error of A, B respectively. To keep the total area at zero, the combined area of C and D has to be zero. Further, for them not to affect the detuned-transition action of the original pulses, the augmenting pulses are required to be $2\pi$-pulses on that transition. It turns out that for first-order $\sigma_z$ error cancellation there are two mutually exclusive conditions (Appendix B 2), namely,

$$|\Omega_C| \tan\frac{\alpha_A}{2} = -|\Omega_A| \tan\frac{\alpha_C}{2}, \tag{21}$$

or

$$\cos\frac{\alpha_A \pm \alpha_C}{2} = 0. \tag{22}$$

Using Eq. (21) together with the zero-area and $2\pi$ requirements yields

$$\Omega_C = -\Omega_D = p \cdot 1.0969\Delta, \tag{23}$$

$$\alpha_C = \alpha_D = 1.478\pi, \tag{24}$$

where $p = \pm 1$. The total duration of the this P-ZAP is

$$T_{P-ZAP} = 4\left(\frac{\alpha_A}{|\Omega_A|} + \frac{\alpha_C}{|\Omega_C|}\right) \approx 3.784 \times \frac{2\pi}{\Delta}. \tag{25}$$

This sequence achieves $\partial EPG/\partial\epsilon = 0$ to all orders and $\partial EPG/\partial\eta = 0$ to first order for the resonant (inactive) transition, while performing an exact $\pi$ rotation on the detuned (active) transition. Its structure and effect on the Bloch spheres are shown in Fig. 2(c). Detuning and power errors on the active transition still contribute at first order, motivating the final construction.

## E. PUDDINGs: Complete Error Protection for Conditional Gates

To protect both transitions against both detuning and power errors, we embed P-ZAPs into a BB-COMP framework, yielding Power-Unaffected, Doubly-Detuning-Insensitive Gates (PUDDINGs). This entails replacing each pulse in a BB-COMP (e.g., U5a) with a P-ZAP, adjusting the overall phase of each P-ZAP according to the BB-COMP prescription. For U5a, this yields

$$U_{PUDDING-U5a} = \prod_{k=1}^{5} e^{-i\phi_k\sigma_z/2} U_{P-ZAP} e^{i\phi_k\sigma_z/2}, \quad (26)$$

where

$$\{\phi_k\} = \{0, 5\pi/6, \pi/3, 5\pi/6, 0\}, \quad (27)$$

and $U_{P-ZAP}$ is the P-ZAP transformation matrix. The total duration of this PUDDING sequence is five times that of the $\pi$ P-ZAP,

$$T_{PUDDING} = 18.92 \times 2\pi/\Delta. \quad (28)$$

The resulting PUDDING inherits the following properties from its consituents: (i) The all-orders power insensitivity on the resonant ('inactive') transition of the ZAP structure; (ii) The first-order detuning insensitivity on the resonant transition of the P-ZAP augmentation; (iii) The first-order power and detuning insensitivity on the detuned ('active') transition of the BB-COMP framework.

Figure 3 compares the Frobenius-distance infidelity of a simple $\pi - 2\pi$ gate [(a),(b)] and a U5a-based PUDDING gate [(c),(d)] as functions of detuning and amplitude errors, for both the inactive [(b),(c)] and active [(a),(d)] subspaces. As is also schematically illustrated in (e) and (f), the narrow valleys of low error surrounding the ideal operating points of the unprotected gates are dramatically broader for PUDDING, clearly demonstrating (at least) quadratic error suppression in both control parameters. The narrow valleys of low error surrounding the ideal operating points of the unprotected gates are dramatically broader for PUDDING, clearly demonstrating (at least) quadratic error suppression in both control parameters.

## F. Effect of dephasing and amplitude noise

Random noise includes inhomogeneous dephasing (quasi-static frequency noise), quasi-static amplitude noise, and homogeneous dephasing (dynamical frequency noise). Inhomogeneous dephasing and quasi-static amplitude noise are captured by averaging the EPG over a detuning and amplitude distribution characterized by $T_2^*$ and $w_\epsilon$, respectively. Homogeneous dephasing (finite $T_2$) is modeled by a time-dependent stochastic detuning with a specified power spectrum (Appendix D). For each noise realization we compute the gate propagator numerically and extract the fidelity relative to the ideal transformation.

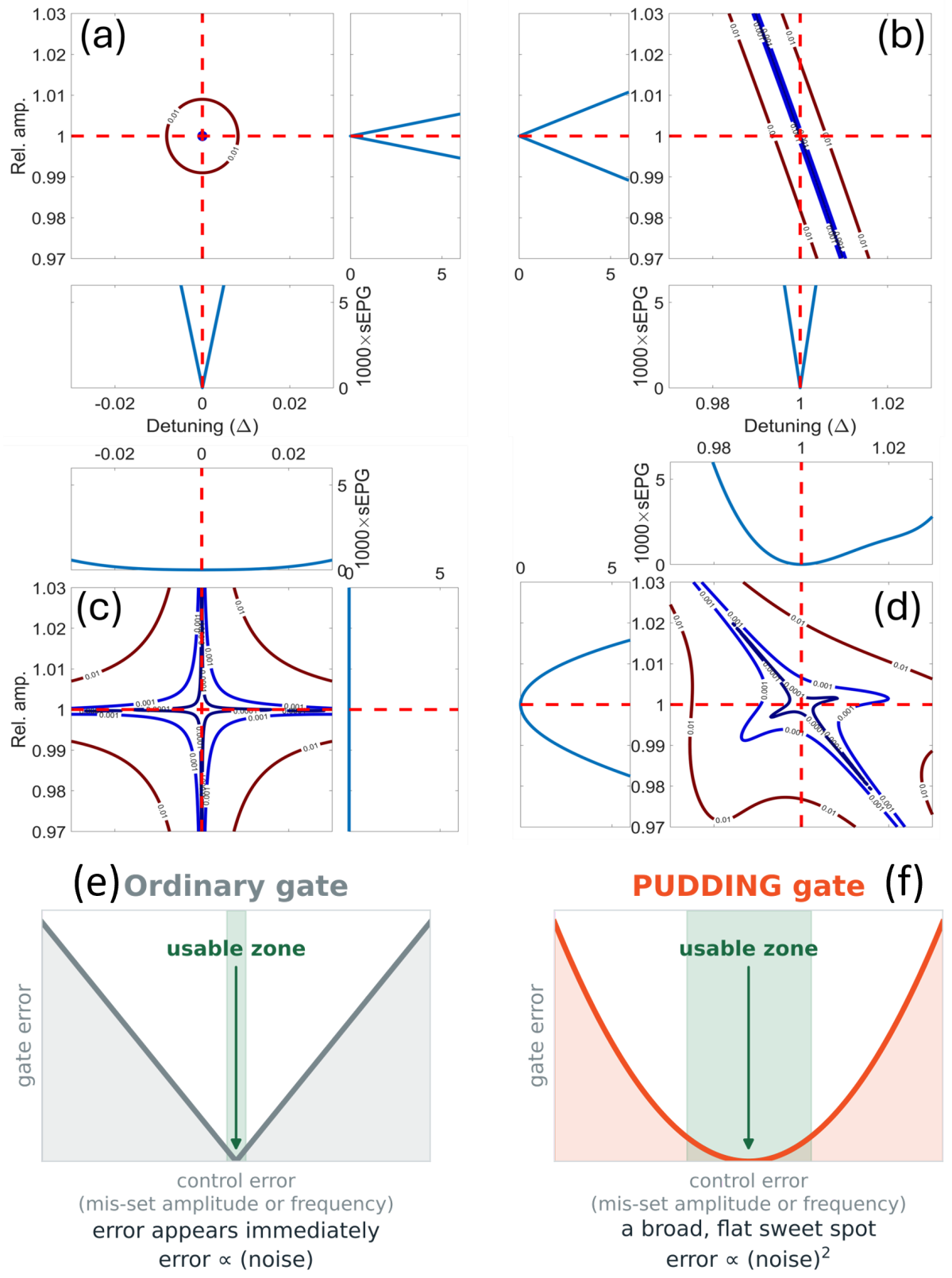

FIG. 3. Two-qubit gate infidelity versus detuning and amplitude: comparison between unprotected and protected gates. (a) and (b) $\pi - 2\pi$ gate. (c) and (d) U5a-based PUDDING gate. The side panels show cross-sections of the fidelity along the dashed red lines in the main panels. (e) [(f)] exemplary cross-sections illustrating the sensitivity of unprotected (protected PUDDING) operation under errors.

Fig. 4 shows the total infidelity for representative single-qubit (a) and two-qubit (b) gates as a function of $T_2^*$, for two values of $w_\epsilon$. For the single qubit gates the time scale is given in terms of the duration of a resonant $\pi$-pulse ($T_\pi = \pi/\Omega$), while for the two-qubit gates it is given in terms of $T_\Delta = 2\pi/\Delta$, the period time related to the coupling frequency $\Delta$. For all gates, the gate time is marked by a vertical dashed line. Fig. 4(b) also includes curves for several finite values of $T_2$ (for the case of no amplitude fluctuations).

As expected, unprotected gates display large infidelities when $T_2^*$ is shorter than the gate duration, with their errors decreasing roughly as $1/T_2^*$ for longer $T_2^*$. In contrast, protected gates remain accurate even when $T_2^*$ is almost an order of magnitude shorter than their (longer) durations, and in the regime where $T_2^*$ exceeds all gate durations their infidelity decreases approximately as $1/T_2^{*2}$, reflecting the first-order detuning insensitivity of the composite design.

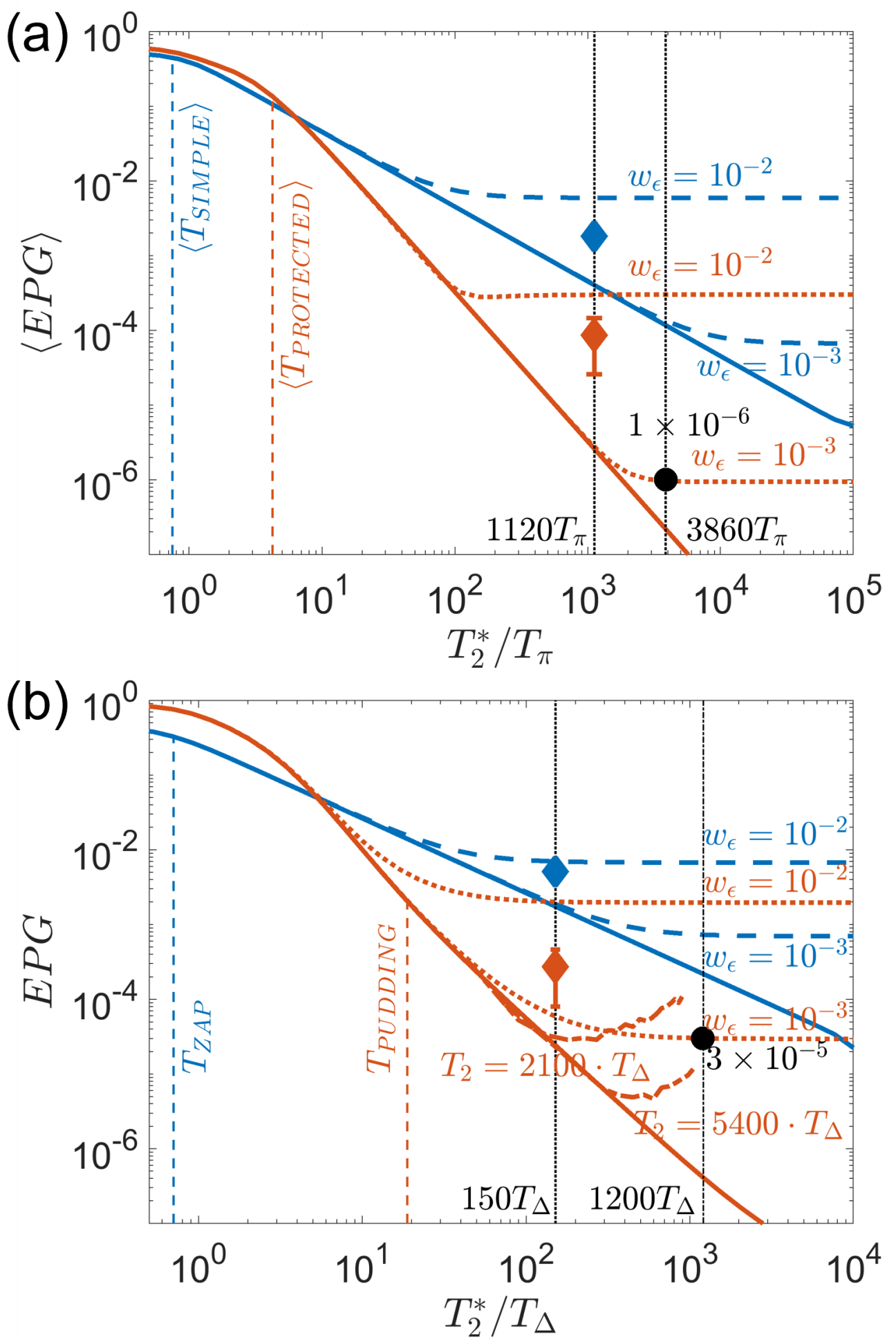

FIG. 4. Gate fidelity versus the inhomogeneous dephasing time. (a) Single-qubit gates. (b) Two-qubit gates. The solid lines assume infinite homogeneous dephasing time and no amplitude fluctuations. The dashed lines include amplitude fluctuations of $10^{-2}$ and $10^{-3}$, as labeled. The symbols and error bars present the experimental results. $T_2^*$ values relevant to the experiments are marked by vertical dotted lines. The Dash-dotted lines in (b) include finite homogeneous dephasing times, as labeled. The prediction for ultra-high isotopic purity diamond at low temperature and amplitude fluctuations of $10^{-3}$ is marked by the black circles.

Both finite amplitude noise and finite homogeneous dephasing impose a floor on achievable fidelity, reached for sufficiently large $T_2^*$. Notably, for both amplitude noise and dynamical frequency noise, while this floor is linear in the noise magnitude for unprotected gate, for protected gates it is quadratic. Thus, in the regime of long-enough $T_2^*$, protected gates always outperform simple gates, the margin increasing with both $T_2^*$ and $1/\sigma_\epsilon$.

## III. EXPERIMENT

Our experiments employ a 20 $\mu$m thick (001) diamond membrane, implanted with $^{15}$N ions at an energy of 60 keV and a density optimized for single-NV isolation. The membrane is made from an isotopically-purified diamond containing at most 0.01% $^{13}$C. The samples were then annealed at 800 °C for 2 hours, leading to the formation of NV centers approximately 30 nm below the surface. Confocal fluorescence microscopy and Hanbury-Brown Twiss intensity correlation measurements ($g^{(2)}(0) < 0.5$) confirmed the location of single NVs. The overall photon collection efficiency, including objective numerical aperture (NA=0.9), optical transmission, and avalanche photodiode (APD) quantum efficiency, reached 8.6%, enabling a high signal-to-noise ratio.

The NV center's electronic ground state is a spin-1 triplet, where the $m_s = 0$ state is split from the $m_s = \pm 1$ states by ∼2.87 GHz at zero magnetic field. A permanent magnet (∼5 mT) oriented along the NV axis further splits the $m_s = \pm 1$ states by ∼300 MHz, isolating $\{|m_s = 0\rangle, |m_s = +1\rangle\}$ as a well-defined qubit. The $^{15}$N nuclear spin (I=1/2) experiences a hyperfine and Zeeman splitting of 2.98 MHz when $m_s = +1$ and 0.05 MHz when $m_s = 0$. This large frequency difference enables selective nuclear spin manipulation conditioned on the electron state, and vice versa.

Electron spin rotations employ resonant microwave (MW) pulses at ∼2.72 GHz, delivered via a printed circuit-board (PCB) loop antenna placed ∼50 $\mu$m beneath the sample. Rabi frequencies $\Omega_e/2\pi \approx 5 - 10$ MHz enable $\pi$ pulses in $\sim 50 - 100$ ns. Nuclear spin rotations utilize radio-frequency (RF) pulses at ∼3 MHz, delivered through the same antenna. Rabi frequencies $\Omega_n/2\pi \approx 30 - 40$ kHz yield $\pi$ pulse durations of $\sim 13 - 17$ $\mu$s. Power stability of the RF amplifier (specified at ±0.5%) and frequency stability ( ±10 Hz) set the instrument-related limits on $\epsilon$ and $\eta$.

A 520 nm laser (100 $\mu$W, 700 ns pulse) spin-polarizes the electron to $|m_s = 0\rangle$ with over 90% efficiency via optical pumping. Fluorescence under 520 nm excitation distinguishes $|m_s = 0\rangle$ (bright) from $|m_s = +1\rangle$ (dark) with a contrast of ∼30%, enabling high-fidelity readout.

A "pure" nuclear spin single-qubit gate (electron spectator) requires compensation for the electron's state-dependent hyperfine shift. This can be done by employing a double-passage protocol: (i) Apply RF pulse at $\omega_n^{(1)}$ (electron in $|{+1}\rangle$). (ii) Flip the electron spin by a MW $\pi$ pulse $|{+1}\rangle \to |0\rangle$. (iii) Apply an RF pulse at $\omega_n^{(1)}$ again. (iv) Flip the electron spin back by a MW $\pi$ pulse $|0\rangle \to |+1\rangle$. Since MW pulses ($\sim$50 ns) are $\sim$100 times faster than RF pulses ($\sim$7 $\mu$s on average), their contribution to gate error is negligible, and the effective single-qubit gate error is dominated by RF pulse imperfections.

We implement a controlled-NOT with the nuclear spin as control and electron as target ($\mathrm{C}_n\mathrm{NOT}_e$) using a zero-area pulse (ZAP) sequence: Apply two consecutive MW pulses resonant with the transition from $|m_s = 0, m_I = -1/2\rangle$ to $|m_s = +1, m_I = -1/2\rangle$. Each pulse has a duration of $\pi/\sqrt{2}\Delta$, where $\Delta \approx 3$ MHz is the MW transition energy difference between states of $m_I = -1/2$ and $m_I = 1/2$. Set the Rabi frequency to $\Delta$ and the relative phase between the two pulses to $\pi$. For $m_I = -1/2$ (resonant), the net rotation is identity (zero area). For $m_I = +1/2$ (detuned by hyperfine), the electron executes a $\pi$ flip. This gate is exact but first-order sensitive; we later test composite versions.

To assess the mean error of single-qubit gates, we implement the Randomized Benchmarking protocol [8]. In this protocol, one first initializes the system to the state $|0_e 0_n\rangle$ by optical pumping, a $\mathrm{C}_n\mathrm{NOT}_e$ gate, an RF $\pi$ pulse, and a second optical pumping. We then bring it to $|1_e 0_n\rangle$ by an $\mathrm{IC}_n\mathrm{NOT}_e$ gate, applied similarly to the $\mathrm{C}_n\mathrm{NOT}_e$ gate, just with the MW resonant with the other transition. After these preparation stages we apply a sequence of $m$ single-qubit Clifford gates $\{C_1, C_2, \ldots, C_m\}$ drawn uniformly from the Clifford group. We implement the full 24-element Clifford group using decompositions into $\{X_{\pi/2}, Y_{\pi/2}, Z_{\pi/2}\}$. Each Clifford is compiled into a sequence of 1 to 3 RF pulses with appropriate phases. After that, we invert the sequence by applying $C_{inv} = (C_m \cdots C_2 C_1)^{-1}$. Finally, we measure the return probability $P_{return}(m)$ of the system to return to the state $|1_e 0_n\rangle$. We apply $N_{trials} > 10^4$ random sequences per $m$, averaging to extract $\langle P_{return}(m)\rangle$.

The return probability decays exponentially: $\langle P_{return}(m)\rangle = Ap^m + B$, where $p$ is the average Clifford survival probability and $A$ and $B$ account for SPAM errors. The EPG is then given by $EPG = (d-1)/d \times (1-p)$, with $d = 2$ for a single-qubit gate. Statistical uncertainties are propagated from the fit covariance matrix. To isolate gate errors from electron relaxation, we perform a reference RB where the RF gates are replaced by idle periods of equal duration. The difference between full RB and reference RB decay rates isolates the RF gate error contribution.

For the two-qubit gates we perform three types of benchamrking. First, we use a simple repeated gate benchmarking protocol (RGB), which consists of state preparation at $|0_e 0_n\rangle$, repeated application of the tested gate for $N/2$ times, followed by $N/2$ applications of the inverse gate, and a measurement of the survival probability of the initial state. While such a protocol is not immune to all systematic errors, and may include long-time interference effects, it is used as an estimator of the difference between the performance of the two tested gates. Next, we perform the RB protocol using conditional rotation (CROT) gates. Finally, we perform the interleaved randomized benchmarking (IRB) protocol, by repeating the RB measurements with the tested CNOT gate interleaved between the random CROT gates. The EPG is then calculated as the difference between the net EPG (after the subtraction of the $T_1$ decay) measured with and without the interleaved CNOT gate. We test both un-protected (ZAP) gates and protected (PUDDING) gates, where in both cases the reference CROT gates for the IRB protocol are PUDDING gates (as they provide a slower-decaying background, increasing the achievable precision).

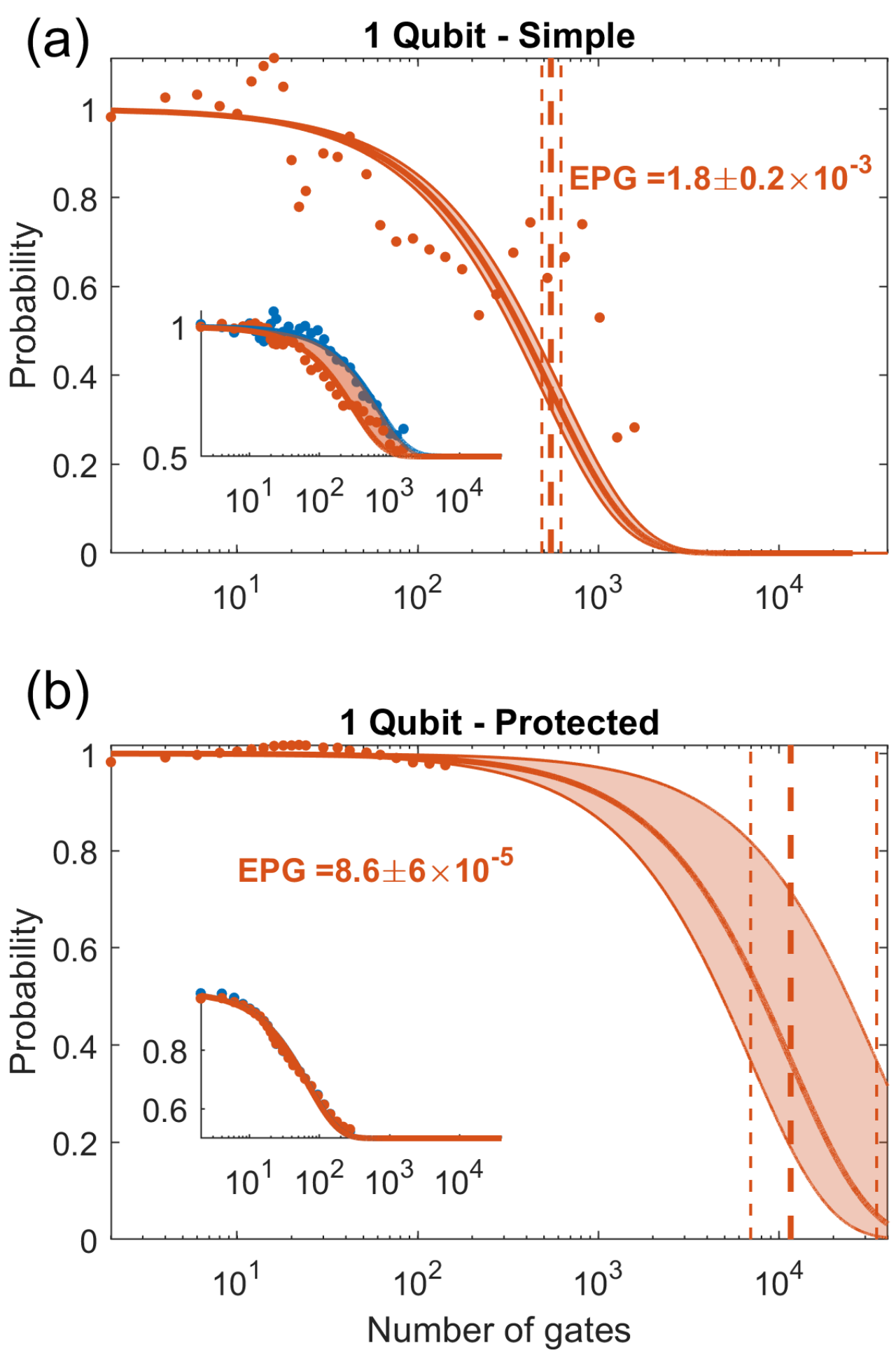

FIG. 5. Experimental benchmarking of single-qubit gates. (a), Randomized benchmarking (RB) of Simple (unprotected) gates. (b) RB of composite (protected) gates. The insets present the raw data, including the reference, electron spin lifetime measurements (in blue).

## IV. RESULTS

### A. Single-qubit gates

We first benchmark nuclear single-qubit gates implemented by simple rectangular RF pulses with an average duration of about 6.5 $\mu$s. Randomized benchmarking yields a net ($T_1$-subtracted) EPG of $1.8 \pm 0.2 \times 10^{-3}$, as presented in Figure 5(a) and in Table I. This error is attributed to RF power fluctuations, nuclear dephasing, and residual cross-talk. Independent estimates from RF amplifier stability and Ramsey measurements indicate quasi-static amplitude and detuning noise at the 0.1–0.2% level and nuclear spin $T_2^*$ of 7.3 ms (measured independently), which together account for errors of a few $10^{-4}$–$10^{-3}$ per gate. These unprotected gates set the baseline performance against which composite protocols are compared.

We next replace each simple RF pulse with a BB-COMPs sequence: the five-pulse U5a structure for $\pi$ rotations and our seven-pulse generalization for arbitrary angles. The average gate duration increases by a factor of about nine. Despite this increase, as can be seen in Figure 5(b) and Table I, randomized benchmarking reveals a substantial improvement: after subtracting the electron-relaxation contribution, we obtain a single-qubit EPG of $8.6 \pm 6 \times 10^{-5}$. Thus, PUDDINGs reduce the nuclear gate error by a factor of about 20.

### B. Two-qubit gates

For two-qubit conditional gates we first benchmark the simple ZAP implementing $C_n NOT_e$. Each of the two MW pulses composing it has duration $\tau \approx 117$ ns, for a total gate time of about 234 ns. As shown in Figure 6(a)-(c), the gate error estimated using RGB, RB, and IRB is $6.5 \pm 0.5 \times 10^{-3}$, $4.4 \pm 0.4 \times 10^{-3}$, and $5.1 \pm 0.2 \times 10^{-3}$, respectively.

We then use the same protocols to test the PUDDING gate [(Figure 6(d)-(f)]. Remarkably, while its total duration is almost 30 times longer than that of the simple ZAP gate, using IRB we measure a $T_1$ subtracted EPG of $2.7 \pm 1.9 \times 10^{-4}$, about 20 times lower than that of the corresponding ZAP EPG.

The measured EPGs for all tested gates are summarized in Table I, and the improvement factors gained are presented in Figure 7.

TABLE I. Summary of Error Per Gate (EPG) results for 1-qubit and 2-qubit gates.

| Error Per Gate (EPG) | | | | |
|---|---|---|---|---|
| | 1 Qubit | 2 Qubit | | |
| Protocol | RB | RGB (CNOT) | RB | IRB (CNOT) |
| ZAP | $(1.8 \pm 0.2) \times 10^{-3}$ | $(6.5 \pm 0.5) \times 10^{-3}$ | $(4.4 \pm 0.4) \times 10^{-3}$ | $(5.1 \pm 0.2) \times 10^{-3}$ |
| PUDDING | $(8.6 \pm 6) \times 10^{-5}$ | $(7.2 \pm 0.4) \times 10^{-4}$ | $(7.4 \pm 1) \times 10^{-4}$ | $(2.7 \pm 1.9) \times 10^{-4}$ |

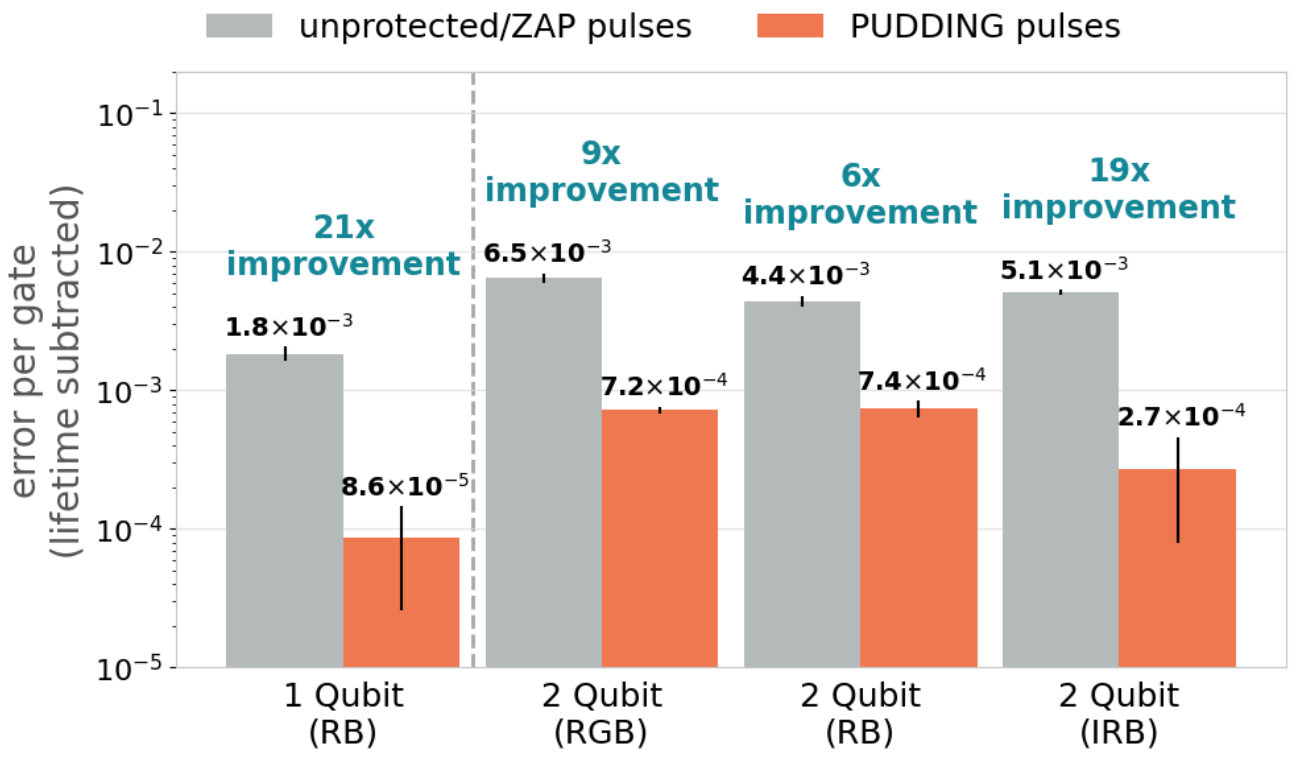

FIG. 7. measured EPG for 1-qubit and 2-qubit gates. visual representation of Table I

## V. DISCUSSION

To compare with our theory, we place the measured EPGs for single qubit gates (RB) and two qubit CNOT gate (IRB) on the $T_2^*$-dependence plots of Figure 4 (diamond symbols).

Using our noise model, the measured room-temperature electron and nuclear spin $T_2$ and $T_2^*$ dephasing times, and the low-temperature, ultra-high purity dephasing times known from the literature (see Table II in Appendix E) we estimate that for a relative amplitude noise of $10^{-3}$ [12],the EPG could be as low as $1 \times 10^{-6}$ and $3 \times 10^{-5}$ for single and two qubit gates, respectively (black dots in Figure 4).

Our results can also be compared with gate infidelities achieved in other leading platforms. Recent benchmarking in superconducting qubits has demonstrated single-qubit EPG at or below $10^{-4}$ and two-qubit EPG around $10^{-3}$ [35]. Trapped-ion systems routinely achieve below-$10^{-4}$ errors for both single- and two-qubit gates, albeit in ultra-high-vacuum and cryogenic environments [36, 37]. Neutral-atom platforms and semiconductor spin qubits are rapidly improving but still under active development [38, 39].

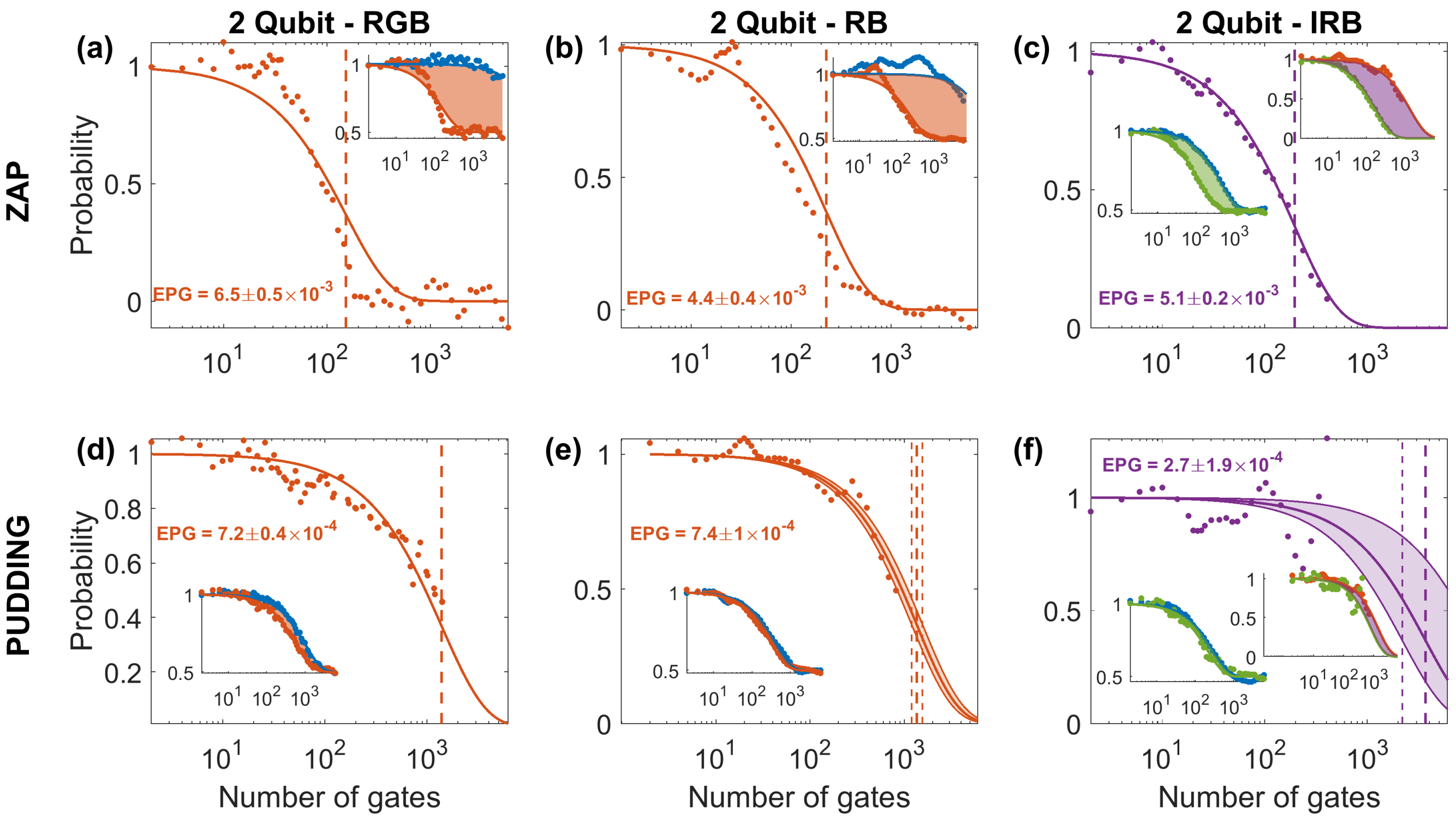

FIG. 6. Experimental benchmarking of two-qubit gates. (a)-(c), Repeated gate benchmarking (RGB), randomized benchmarking (RB), and interleaved randomized benchmarking (IRB) of ZAP gates, respectively. (d)-(f) same as (a)-(c) for PUDDING gates. The measured population-decay-normalized survival probability (model fit) is presented versus the total number of gates by circles (solid lines). The inset presents the raw data, where the population decay is presented in blue. The vertical dashed lines present the $1/e$ gate number for each measurement. For IRB [(c), (f)], the raw survival probability data and fit are presented in the lower-left inset by the circles and line, together with the population decay measurement and fit (in blue). The upper-right inset presents the population-decay-normalized survival probability (model fit) by circles (solid line), together with the population-decay-normalized PUDDING RB survival probability [taken from main panel (e)].

In our room-temperature NV experiment with natural-abundance diamond, PUDDING-protected single-qubit gates achieve $EPG < 1 \times 10^{-4}$, comparable to state-of-the-art superconducting-qubit performance, but in a solid-state defect platform operating at ambient conditions. As shown in Figure 8, two-qubit PUDDING gates are on-par with the best superconducting and ion-trap results, with $EPG \approx 2 - 7 \times 10^{-4}$, an order-of-magnitude improvement over previous NV two-qubit gates [4, 13]. When projected to ultra-pure, isotopically enriched $^{12}$C diamond at 4K, the PUDDING protocol approaches or surpasses the two-qubit error levels reported in other platforms, while retaining the unique advantages of NV centers for integrated quantum networking.

Beyond the specific NV implementation, the PUDDING construction addresses a broader theoretical question: can conditional gates be made intrinsically robust to both amplitude and detuning errors on both the identity and flip branches without sacrificing exactness? Our answer is yes, via a combination of zero-area structure, augmented symmetry, and composite embedding.

Further, because the only assumptions are a driven two-level system with a conditionally shifted transition frequency and the ability to shape amplitude and phase of the drive, PUDDINGs can be implemented in any architecture with energy-selective conditional interactions. This includes superconducting transmons with tunable couplers [40], exchange-coupled spins in semiconductor quantum dots [39, 41], trapped ions with state-dependent AC Stark shifts [28, 42], Rydberg-based neutral atoms [38], and rare-earth ion systems with hyperfine-coupled electron–nuclear pairs, and could thus contribute across the field of quantum computing.

## VI. CONCLUSIONS

We have introduced and experimentally validated a comprehensive framework for error-protected quantum gates in NV centers in diamond, bridging theoretical pulse design with room-temperature randomized benchmarking. Our main achievements are: (i) experimental benchmarking of single- and two-qubit gates at room temperature, including a detailed error budget that isolates contributions from electron relaxation, dephasing, and control-power fluctuations; (ii) the development of PUDDINGs, a new class of composite conditional gates that combine protected zero-area pulses with broadband composite sequences to achieve first-order insensitivity to both amplitude and detuning errors on both active and inactive subspaces; and (iii) demonstration of substantial error suppression, with PUDDING-protected single-qubit gates achieving $EPG < 1 \times 10^{-4}$ and two-qubit gates achieving $EPG \approx 3 - 7 \times 10^{-4}$.

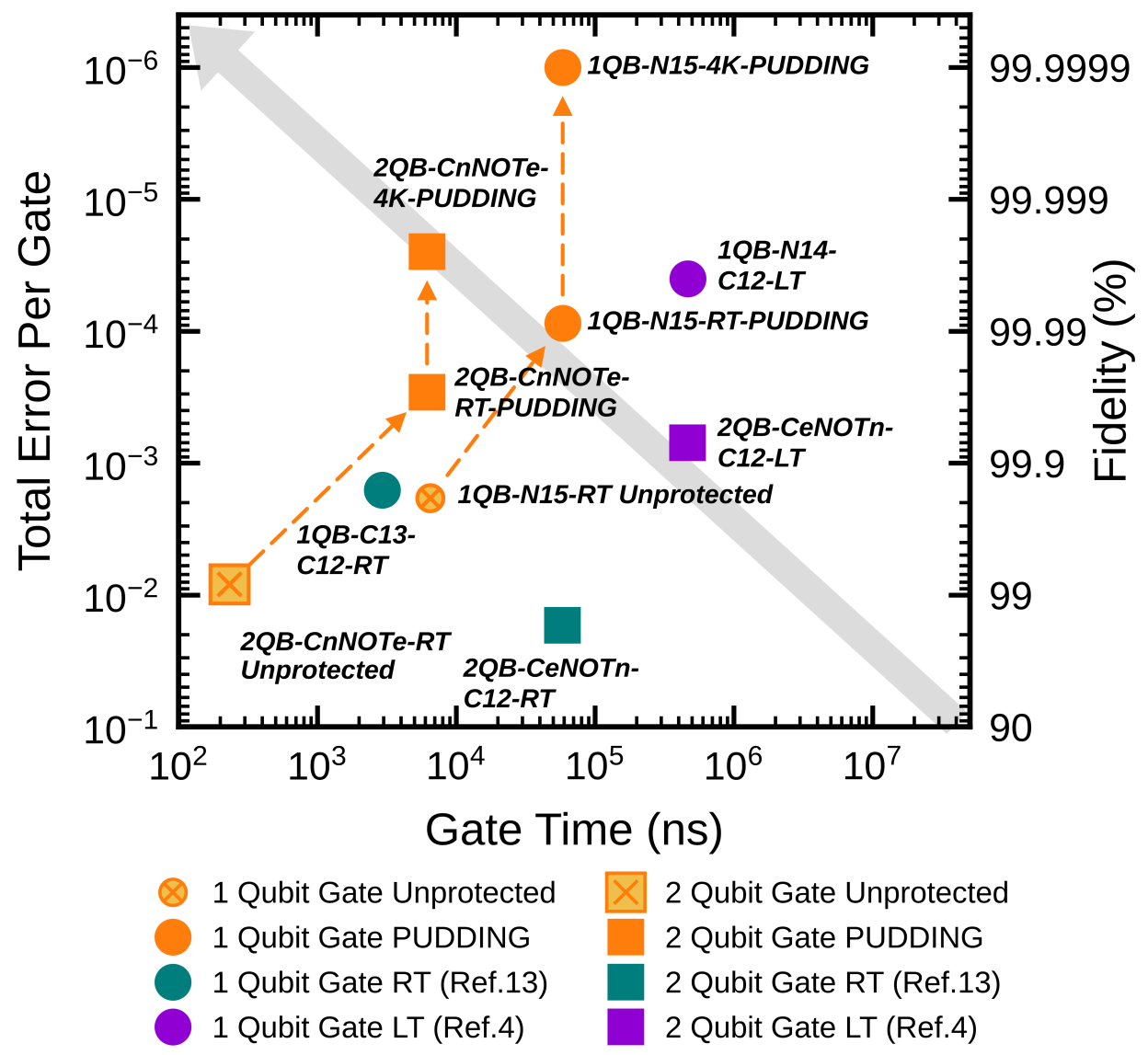

FIG. 8. (a) Error per gate versus gate time of achieved in state-of-the-art experiments using NV centers. Yellow and orange symbols - this work. Green symbols - room temperature (RT) results from Ref. 13. Purple symbols - low temperature (LT) results from Ref. 4.

Using experimentally informed noise models, we project that in isotopically purified $^{12}$C diamond operated at cryogenic temperatures, single-qubit protected gates can achieve an error per gate as low as $1 \times 10^{-6}$, while their two-qubit PUDDING counterparts can achieve $3 \times 10^{-5}$, more than two orders of magnitude below standard surface- and color-code thresholds. These results demonstrate that NV centers can support gate fidelities competitive with leading superconducting [35, 43] and trapped-ion [36, 37] platforms while offering unique advantages for quantum networking, including optical connectivity and long-lived nuclear memories.

## Appendix A: Derivation of single qubit protected arbitrary gate

The rules of the game: We are given rotation matrices of two types. One type rotates by $\pi$, and the other rotates by some angle $\beta$. All the matrices rotate around an axis in the $x$-$y$ plane. For each matrix, we are allowed to choose the phase, that is, the angle, $\phi$, of the rotation axis from the $x$ axis. We are then asked to compose these matrices in such a way that the composite matrix would:

- Rotate by a given angle, $\alpha$ about an axis in the $x$-$y$ plane.
- Be insensitive, to first order, to amplitude errors common to all the composing matrices.
- Be insensitive, to first order, to frequency errors common to all the composing matrices.

First, we have to determine the minimal number of matrices that are required. This can be done by counting the number of parameters needed to be set and the number of available conditions. Let's assume that we take $\pi$ matrices of $n$ different phases (so we take at least $n$ $\pi$ matrices, but can take more), and $\beta$ matrices of $m$ different phases (so we take at least $m$ such matrices but can take more). Note that $m$ cannot be zero, as then we would be limited to $\pi$ or $2\pi$ rotations only, and we want to have an arbitrary overall rotation angle. As a global phase does not matter, we thus have at most $m + n - 1$ phases to set. Further, we have to set $\beta$. So, in total, we'll have at most $m + n$ parameters to set.

How many conditions do we have? Any rotation matrix has three parameters in total. Of these, one parameter is left free (as we're not restricted to a specific angle in the $x$-$y$ plane of the overall rotation axis), so this gives two conditions. Further, we want the first derivatives with respect to both amplitude and frequency deviations of all three parameters to be zero. This gives six additional conditions. Therefore, we have eight conditions in total. So, $m + n = 8$.

Now, if we work with time-symmetric sequences, the amplitude and frequency derivatives of two out of the three parameters of the composed matrix are automatically set to zero. This leaves only four conditions and therefore simplifies the problem considerably. So, we now have $m + n = 4$. As the pulse sequence is symmetric, it will have to contain at least $2(m + n) - 1 = 7$ pulses.

Without loss of generality, we choose $m = 1$, $n = 3$, and place the two $\beta$ pulses at the beginning and the end of the sequence. This gives a sequence of the form $\beta_0 \pi_{\phi_1} \pi_{\phi_2} \pi_{\phi_3} \pi_{\phi_2} \pi_{\phi_1} \beta_0$, where, for example, $\pi_{\phi_1}$ indicates a $\pi$ rotation about an $x$-$y$ plane axis at an angle $\phi_1$ from the $x$ axis.

Having set the form of the sequence, finding the equations for the four conditions and solving them for the four parameters, $\beta, \phi_1, \phi_2$, and $\phi_3$, reduces to a matter of some (rather lengthy but not too complicated) algebra. We obtain:

$$\begin{aligned}\beta &= \frac{\pi - \alpha}{2},\\ \phi_1 &= g(\alpha),\\ \phi_2 &= 2g(\alpha) + h(\alpha),\\ \phi_3 &= 2g(\alpha) + 2h(\alpha),\end{aligned}$$

with

$$\begin{aligned}g(\alpha) &= \cos^{-1}\left(-\frac{\alpha}{4\pi} - \frac{1}{4}\cdot\sin\frac{\alpha}{2}\right),\\ h(\alpha) &= \cos^{-1}\left(-\frac{\alpha}{4\pi} + \frac{1}{4}\cdot\sin\frac{\alpha}{2}\right).\end{aligned}$$

Finally, having found the parameters of the rotation matrices, translating them into pulse parameters is straightforward. For example, for the most simple case of a constant Rabi frequency, $\Omega$, the time durations of the pulses in the sequence would be simply $\pi/\Omega$ and $\beta/\Omega$ for the $\pi$ and $\beta$ matrices, respectively. The phase of each pulse would just be the rotation axis angle of the corresponding rotation matrix.

## Appendix B: Conditions for detuning insensitivity near resonance

### 1. $\sigma_x$ and $\sigma_y$: symmetric zero-$\pi$ pulses

For a near-resonant pulse with a time-dependent, real Rabi frequency, $\Omega(t)$, and a fixed, small detuning, $\delta$, the Hamiltonian is

$$H = \frac{\Omega(t)}{2}\sigma_x + \frac{\delta}{2}\sigma_z. \tag{B1}$$

Performing a Magnus expansion up to second order in $H$, the propagation operator for the entire pulse (from $t=-\infty$ to $t=\infty$) is approximately

$$U \approx e^{M_1+M_2}, \tag{B2}$$

where,

$$M_1 = -i\int_{-\infty}^{\infty} H(t)dt, \tag{B3}$$

$$M_2 = -\frac{1}{2}\int_{-\infty}^{\infty} dt \int_{-\infty}^{t} \left[H(t), H(t')\right]dt'. \tag{B4}$$

Inserting Eq. (B1) into Eq. (B3) and Eq. (B4) we obtain

$$M_1 = \frac{i}{2}\sigma_x\int_{-\infty}^{\infty}\Omega(t)dt + \frac{i}{2}\sigma_z\delta T + O(\delta^2), \tag{B5}$$

$$M_2 = \frac{i}{8}\delta\sigma_y\left(\int_{-\infty}^{\infty} t\Omega(t)dt - \int_{-\infty}^{\infty}\int_{-\infty}^{t}\Omega(t')dt'dt\right) + O(\delta^2), \tag{B6}$$

where $T$ is the pulse duration (for a finite-support pulse profile).

From Eq. (B5) we see that for $\Omega(t)$ whose area is zero (a zero-$\pi$ pulse), the $\sigma_x$ component vanishes up to second order in $\delta$. From Eq. (B6) we see that for $\Omega(t)$ which is both a zero-$\pi$ pulse and is symmetric in time ( $\Omega(-t) = \Omega(t)$), the $\sigma_y$ component also becomes first-order insensitive to $\delta$, as for this case both terms on the right-hand side of Eq. (B6) vanish. The first term vanishes as it is the first moment of an even function. The second term vanishes as the integral of the Rabi frequency, which is even and starts at zero for $t=-\infty$, is an odd function, and thus its total area is also zero.

In a similar way, one can also show that for a symmetric zero-$\pi$ pulse, all higher orders of the expansion contribute only $\sigma_x$ and $\sigma_y$ terms that are at least quadratic in $\delta$. However, from Eq. (B5) we see that having a symmetric zero-$\pi$ pulse is not a sufficient condition for the $\sigma_z$ term to become first-order insensitive to $\delta$, and thus additional conditions are required for complete first-order detuning protection.

### 2. $\sigma_z$: augmented pulses

In order to compensate also for $\sigma_z$ errors in the 'inactive' (on-resonance) sector of a symmetric ZAP ( $ABBA$), we augment each of the four pulses with an extra pulse whose parameters are chosen to provide the desired first-order cancellation. As pulse $B$ is just the negative of pulse $A$, we can consider only the augmentation pulse of the latter, which we call pulse $C$. Pulse $D$, augmenting pulse $B$, will be the negative of pulse $C$. For pulses $A$ and $C$, the Rabi frequencies are $\Omega_A$ and $\Omega_C$ and the rotation angles $\alpha_A$ and $\alpha_C$, respectively. The transition matrices, up to first order in the detuning $\delta$ are,

$$\begin{aligned}U_{A(C)} &= \cos\frac{\alpha_{A(C)}}{2}I\\ &- ip_{A(C)}\sin\frac{\alpha_{A(C)}}{2}\sigma_x - i\sin\frac{\alpha_{A(C)}}{2}\frac{\delta}{|\Omega_{A(C)}|}\sigma_z,\end{aligned} \tag{B7}$$

where $p_{A(C)} = \text{sgn}(\Omega_{A(C)})$ is the sign of the Rabi frequency.

The transition matrix of the combined pulse, $CA$, up to first order in $\delta$ is then

$$\begin{aligned}U_{CA} &= \cos\left(\frac{\alpha_A + p\alpha_C}{2}\right) I - ip_A \sin\left(\frac{\alpha_A + p\alpha_C}{2}\right)\sigma_x \\ &- ip \sin\frac{\alpha_A}{2}\sin\frac{\alpha_C}{2}\left(\frac{1}{\Omega_C} - \frac{1}{\Omega_A}\right)\delta\sigma_y \\ &- i\left(\frac{1}{|\Omega_A|}\sin\frac{\alpha_A}{2}\cos\frac{\alpha_C}{2} + \frac{1}{|\Omega_C|}\cos\frac{\alpha_A}{2}\sin\frac{\alpha_C}{2}\right)\delta\sigma_z,\end{aligned} \tag{B8}$$

where $p = p_A p_C$ is the relative sign between $\Omega_C$ and $\Omega_A$.

We therefore see that the $\sigma_z$ component of the combined pulse $CA$ is insensitive to $\delta$ up to first order if

$$|\Omega_C| \tan\frac{\alpha_A}{2} = -|\Omega_A| \tan\frac{\alpha_C}{2}, \tag{B9}$$

which proves Eq. (21). It is straightforward to see that the same condition applies also for $AC$.

To see that this condition is enough for the entire pulse to be first-order insensitive to the detuning, we note that under this condition, the pulse $DB$ has the same transition matrix as that of $CA$, up to an opposite sign for the $\sigma_x$ and $\sigma_y$ components. This is because $\alpha_B = \alpha_A$, $\Omega_B = -\Omega_A$, and the relative sign between $\Omega_D$ and $\Omega_B$ is the same as that between $\Omega_C$ and $\Omega_A$. Thus, for $DBCA$, these components will vanish, and it will have no first-order detuning dependency. In turn, this means the same for the full composite pulse $ACBDDBCA$.

For the case that Eq. (21) does not hold, one can show using similar arguments that the transition matrix of the four component pulse $DBCA$ would read, up to first order in $\delta$,

$$\begin{aligned}U_{DBCA} = I - 2i\eta\delta\Bigg[&\sin\left(\frac{\alpha_A + p\alpha_C}{2}\right)\sigma_y \\ &+ \cos\left(\frac{\alpha_A + p\alpha_C}{2}\right)\sigma_z\Bigg],\end{aligned} \tag{B10}$$

with $\eta = \left(\frac{1}{|\Omega_A|}\sin\frac{\alpha_A}{2}\cos\frac{\alpha_C}{2} + \frac{1}{|\Omega_C|}\cos\frac{\alpha_A}{2}\sin\frac{\alpha_C}{2}\right)$.

One can show that the transition matrix of the second part of the full pulse, $U_{ACBD}$, would have the same structure, except for an opposite sign for the $\sigma_y$ component. Thus, the full transition matrix would be of the form

$$\begin{aligned}U_{ACBDDBCA} &= I \\ &- 4i\cos\left(\frac{\alpha_A + p\alpha_C}{2}\right)\eta\delta\sigma_z + O(\delta^2).\end{aligned} \tag{B11}$$

This shows that in addition to the condition of Eq. (21) there is an alternative condition for first-order insensitivity of the full augmented symmetric zero-area pulse to $\delta$:

$$\cos\left(\frac{\alpha_A + p\alpha_C}{2}\right) = 0, \tag{B12}$$

which proves Eq. (22).

## Appendix C: Symmetric zero-$\pi$ pulses of arbitrary conditional rotation angle

We consider an operation of the form,

$$U = U_+ U_- U_- U_+, \tag{C1}$$

where

$$U_\pm = \cos\frac{\tilde{\alpha}}{2} I - i\sin\frac{\tilde{\alpha}}{2}(\pm\sin\beta\sigma_x + \cos\beta\sigma_z). \tag{C2}$$

Here

$$\tan\beta = \frac{\Omega}{\Delta}, \tag{C3}$$

where $\Omega$ is the Rabi frequency and $\Delta$ is the detuning of the off-resonant transition, and $\tilde{\alpha}$ is the rotation angle of the detuned transition. The rotation angle of the resonant transition, $\alpha$, is related to $\tilde{\alpha}$ by

$$\alpha = \tilde{\alpha} \cdot \cos\beta. \tag{C4}$$

This form of $U$ ensures that the total pulse area vanishes and that the pulse sequence is time-symmetric. While it is not the most general form to do so, it is one of the simplest, as it requires only a single amplitude value for all the components.

As we have two parameters to set, $\tilde{\alpha}$ and $\beta$, we need two requirements. The first is that the total rotation induced by $U$ for the detuning $\Delta$ has no $z$ component, such that for the detuned transition, $U$ acts as if it was a simple, resonant pulse. Thus, the coefficient of $\sigma_z$ within $U$ must vanish. One can show that this leads to the equation

$$\cos\tilde{\alpha}\cos^2\beta + \sin^2\beta = 0. \tag{C5}$$

The second requirement is that the overall rotation angle induced by $U$ for the detuned transition has a specified value, $\theta$. This means that the coefficient of the $I$ component of $U$ is $\cos(\theta/2)$. This leads to the equation

$$\cos 2\tilde{\alpha}\cos^2\beta + \sin^2\beta = -\cos\frac{\theta}{2}. \tag{C6}$$

One can then reach an equation for $\tilde{\alpha}$,

$$\cos\tilde{\alpha}(\cos 2\tilde{\alpha} - 1) = (\cos\tilde{\alpha} - 1)\left(\cos 2\tilde{\alpha} + \cos\frac{\theta}{2}\right). \tag{C7}$$

Expanding $\cos 2\tilde{\alpha}$ in terms of $\cos\tilde{\alpha}$, dividing both sides by $\cos\tilde{\alpha} - 1$ ($\tilde{\alpha} \neq 0$), and rearranging, we arrive at

$$\cos\tilde{\alpha} = -\sin^2\frac{\theta}{4}. \tag{C8}$$

Plugging this into Eq. (C5) then yields

$$\tan\beta = \sin\frac{\theta}{4}, \tag{C9}$$

which together with Eq. (C3) gives Eq. (17). Finally, using Eq. (C4) we obtain

$$\alpha = \cos\beta \cos^{-1}\left(-\sin^2\frac{\theta}{4}\right), \tag{C10}$$

which together with Eq. (C9) gives Eq. (18).

## Appendix D: Simulation of dynamical noise

In order to add a finite homogeneous dephasing time to our model, we introduce a random, time-varying detuning, $\delta(t)$, to the Hamiltonian and solve for the propagation matrix numerically. We take the frequency content of $\delta(t)$ to be given by a truncated $1/f$ distribution:

$$\delta(t) = \frac{A}{T_2^*}\sum_{n=1}^{N}\frac{1}{n}\cos\left(\frac{2\pi f_c}{N}nt + \phi_n\right), \tag{D1}$$

where $\phi_n$ is a random number drawn from a uniform distribution between 0 and $2\pi$, $f_c$ is the cutoff frequency, $N$ is fixed number much larger than 1 (we take $N = 100$), $T_2^*$ is the inhomogeneous dephasing time, and $A$ is a dimensionless constant. We first set $1/f_c \gg T_2^*$ and fix $A$ such that the decay time of a simulated Ramsay experiment is $T_2^*$. Then, for each value of $T_2^*$, we set $f_c$ such that the decay time of a simulated Hahn echo experiment is $T_2$.

We repeat the calculation for fifty realizations of $\delta(t)$. For each realization we calculate the mean of the squares of the differences between the elements of the obtained matrix and the ideal matrix, and take the over-all fidelity to be the square root of the mean of all of theses values.

## Appendix E: Time scales

Table II presents the lifetimes and coherence times measured for our sample, together with the state of the art.

| Spin | | 300K | 4K |
|---|---|---|---|
| **Electron** this work | $T_1$ (sec) | $6.26 \times 10^{-3}$ | |
| | $T_2$ (sec) | $7.36 \times 10^{-4}$ | |
| | $T_2^*$ (sec) | $5 \times 10^{-5}$ | |
| **Electron** State of the art | $T_1$ (sec) | $7.4 \times 10^{-3}$ [44] | $> 3.6 \times 10^{3}$ [19] |
| | $T_2$ (sec) | $2.43 \times 10^{-3}$ [44] | $6.8 \times 10^{-3}$ [11] |
| | $T_2^*$ (sec) | $1.54 \times 10^{-3}$ [44] | $4 \times 10^{-4}$ [11] |
| **Nuclear** State of the art | $T_1$ (sec) | 260 [31] | $> 360$ [45] |
| | $T_2$ (sec) | $8.6 \times 10^{-3}$ [31] | 2.3 [45] |
| | $T_2^*$ (sec) | $7.25 \times 10^{-3}$ [46] | $2.51 \times 10^{-2}$ [45] |

TABLE II. Measured lifetimes and coherence times.